\documentclass[3p, 12pt]{elsarticle}

\usepackage{graphicx,amscd,amsmath,amssymb,verbatim}
\usepackage{amsfonts,epsfig}
\usepackage{mathptmx}
\usepackage{setspace}
\usepackage{epsfig}
\usepackage{url}
\usepackage{multirow}

\begin{document}

\journal{Elsevier}

\begin{frontmatter}

\title{A Simple Non-Markovian Computational Model of the Statistics of Soccer 
Leagues: Emergence and Scaling effects}

\author{Roberto da Silva$^{1}$}
\author{Mendeli Vainstein$^{1 }$}
\author{Luis Lamb$^{2 }$} 
\author{Sandra Prado$^{1}$ }

\address{1 - Instituto de F\'{\i}sica\\
Universidade Federal do Rio Grande do Sul\\
Av. Bento Gon\c{c}alves 9500 \\ 
Caixa Postal 15051
 91501-970, Porto Alegre RS, Brazil\\
%Fax: +55 51 3308 7308\\
%Tel: +55 51 3308 9464\\
{\normalsize {E-mail:{rdasilva,mendeli,sprado}@if.ufrgs.br}}}

\address{2 - Instituto de Inform\'atica\\
Universidade Federal do Rio Grande do Sul\\
Av. Bento Gon\c{c}alves 9500 \\ 
Caixa Postal 15064
91501-970, Porto Alegre, RS,  Brazil\\
%Fax: +55 51 3308 7308\\
%Tel: +55 51 3308 9464\\
{\normalsize{E-mail:lamb@inf.ufrgs.br}}}

%\keywords{1 - Statistics in Sports, 2 - Evolutionary automata for soccer
%tournament modeling, 3 - Analysis of fluctuations resulting from variation
%in the number of teams in tournaments}

\begin{abstract}
{
We propose a novel algorithm that outputs the final standings of a soccer
league, based on a simple dynamics that mimics a soccer tournament. In our
model, a team is created with a defined potential(ability) which is updated
during the tournament according to the results of previous games. The
updated potential modifies a teams' future winning/losing probabilities. We
show that this evolutionary game is able to reproduce the statistical
properties of final standings of actual editions of the Brazilian tournament
(\textit{Brasileir\~{a}o}). However, other leagues such as the Italian
(Calcio) and the Spanish (La Liga) tournaments have notoriously non-Gaussian
traces and cannot be straightforwardly reproduced by this evolutionary
non-Markovian model. A complete understanding of these phenomena deserves
much more attention, but we suggest a simple explanation based on data
collected in Brazil: Here several teams were crowned champion in previous
editions corroborating that the champion typically emerges from random
fluctuations that partly preserves the gaussian traces during the
tournament. On the other hand, in the Italian and Spanish leagues only a few
teams in recent history have won their league tournaments. These leagues are
based on more robust and hierarchical structures established even before the
beginning of the tournament. For the sake of completeness, we also elaborate
a totally Gaussian model (which equalizes the winning, drawing, and losing
probabilities) and we show that the scores of the
Brazilian tournament \textquotedblleft Brasileir\~{a}o" cannot be
reproduced. This shows that the evolutionary aspects are not
superfluous in our modeling and have an important role, which must be
considered in other alternative models. Finally, we analyse the distortions
of our model in situations where a large number of teams is considered,
showing the existence of a transition from a single to a double peaked histogram
of the final classification scores. An interesting scaling is presented for different sized 
tournaments.
}
\end{abstract}

\end{frontmatter}

\setlength{\baselineskip}{0.7cm}

\section{Introduction}

Soccer is an extremely popular and profitable, multi-billion dollar business
around the world. Recently, several aspects regarding the sport and
associated businesses have been the subject of investigation by the
scientific community, including physicists who have devoted some work and
time to describe statistics related to soccer. In the literature about
soccer models, one can find applications of complex networks~\cite{Onody2004}
and fits with generalized functions~\cite{Renio2000}; however, they ofttimes have 
only one focus: goal distribution (see e.g. ~\cite {Bitner2007, Bitner2009, Skinera2009}). 
Outside the soccer literature, it is important to mention other interesting studies 
which do not necessarily focus on the scores of the games, such as models that 
investigate properties of patterns emerging from failure/success processes in sports.
 In the case of basketball, it has been suggested~\cite{yaari2011} that the 
``hot hand" phenomenon (the belief that during a particular period a player's
performance is significantly better than expected on the basis of a player's
overall record), a definitively a non-random pattern, can be modeled by a
sequence of random independent trials. Returning to soccer, some 
authors~\cite{Kranjec2010} have devoted attention to the influence of the
perceptual-motor bias associated with reading direction in foul judgment by
referees.

However, it is interesting to notice that there is a void in the literature:
 few studies have been carried out under the game theoretic approach
of considering the outcome of a tournament from a simple dynamics among the
competing teams. In other words, in looking at the statistics that emerge
from this complex system called soccer, one can ask if the properties of the
distribution of final tournament classification points can be seen as an
emerging property of a soccer tournament dynamics established by simple
rules among the different competing teams, or how these classification point
distributions emerge from a soccer tournament by considering all
\textquotedblleft combats" among the teams. Here, we propose a model that
combines previous studies concerning goal distribution~\cite{Skinera2009}
and a game theoretic approach to football tournaments that produces
realistic final tournament scores and standings.

In this paper, we explore the statistics of standing points in the end of
tournaments disputed according to the ``Double Round Robin System" (DRRS)%
\footnote[1]{%
http://en.wikipedia.org/wiki/Round-robin\_tournament} in which the team with
the most tournament points at the end of the season is crowned the champion,
since many soccer tournament tables around the world are based on this
well-known system. In general, 20 teams take part in the first tier
tournament, such as ``Serie A'' in Italy, the English ``Premier League'',
the Spanish ``La Liga'' and the Brazilian ``Brasileir\~{a}o'' (from 2003
onwards) soccer tournaments. During the course of a season, each team plays
every other team twice: the ``home'' and ``away'' games. Moreover the points
awarded in each match follows the 3-1-0 points system: teams receive three
points for a win and one point for a draw; no points are awarded for a loss.
The Serie A Italian soccer tournament, or simply the ``Calcio", has been
played since 1898, but only from 1929 was it disputed in its current format
and system. Their main champions have been Juventus, winner of the league 27 times, and 
Milan and Internazionale which won the league 18 times each. The Spanish
``La Liga" also started in 1929, and over its history, the tournament has
been widely dominated by only two teams: Real Madrid and Barcelona.

In Brazil, the national tournament, popularly known as ``Brasileir\~{a}o",
was first organized in a modern format in 1971. In 2010 the Brazilian Soccer
Confederation (CBF) recognized as national champions the winners of smaller
national tournaments such as the ``Ta\c{c}a Brasil" (played from 1959 to
1968) and another tournament known as ``Roberto Gomes Pedrosa" (played from
1967 to 1970). However, only in 2003 the Brazilian League started being
disputed via the DRRS. In all past editions of the tournament the league
table was based on the method of preliminaries, typically used in Tennis
tournaments, which will not be considered in this paper. In the 10 editions
played under the DRRS, the brazilian tournament has already been won by 6
different football clubs: Cruzeiro, Santos, S\~{a}o Paulo, Corinthians,
Flamengo, and Fluminense.

The statistics as well as the fluctuations associated to the standings and
scores of teams in tournaments with 20 teams playing under the DRRS can be
very interesting. Moreover, if we are able to reproduce such statistics via
a simple automaton considering the teams as ``agents" which evolve according
to definite ``rules" based on their previous performances and conditions,
one could use this information when preparing or building up a team before a
competition. Thus, models (e.g. automata) of games in a tournament, whose
results are defined by the evolving characteristics of the teams, could
provide important knowledge. Therefore, by exploring the conditions under
which the standing and scores of tournaments can be mimicked by a model, we
propose a simple, but very illustrative, evolutionary non-Markovian process.

It is known that many events can alter the performance of teams during a
season besides their initial strengths, such as the hiring of a new player,
renewed motivation due to a change in coach, key player injuries, trading of
players, among others. For the sake of simplicity, we consider that the
teams in the model initially have the same chance of  winning the games and that the
combination of events that can lead to an improvement of a team will be
modeled solely by increasing the probability of a team winning future games
after a victory. Similarly, a loss should negatively affect their future winning
probabilities.

Our main goal is to verify if the Brazilian Soccer tournament has final
standing scores with the same statistical properties that emerge from our
simple model, and to check whether the properties of the Brazilian
tournament differ from other leagues and, if so, the reasons for that
behavior. In the first part of the paper we calibrate our model by using
constant draw probabilities introduced ad hoc, based on data from real
tournaments. In the second part, we have used draw probabilities that emerge
from the model dynamics, being dependent on the teams \textquotedblleft
abilities". Both situations are able to reproduce real tournament data.
The advantage of the second approach is the independence of extra 
parameters, i.e., the first one uses pre-calculated rates from previous
statistics. In addition, we analyze distortions of our model under hypotheses of
inflated tournaments. Finally, we show a
transition from single to double peaked histograms of final standing
scores, which occurs when we analyze a small league and large tournaments. 
However, it is possible to obtain a scaling for different
tournaments with different sizes.

\section{A first Model: ad-hoc draw probabilities}

In our model, each team starts with a potential $\varphi _{i}(0)=\varphi_0$, where 
$i=1,..,n$ indexes the teams. Each team plays once
with the other $n-1$ teams in each half of the tournament; 
%. The tournament is divided in two halves, so 
a team $A$ plays with $B$ in the first half of the tournament and $B$ plays
with $A$ in the second, i.e. the same game occurs twice in the tournament
and there is no distinction between home and away matches (the
\textquotedblleft home court advantage\textquotedblright\ could be inserted 
in the potential of the teams). In a game between team $i$ and team 
$j$, the probability that $i$ beats $j$ is given by 
\begin{equation}
\Pr (i\succ j)=\frac{\varphi _{i}}{(\varphi _{i}+\varphi _{j})}.
\label{prob_win}
\end{equation}

The number of games in the tournament is $N=n(n-1)$ and in each half of the
tournament, $n-1$ rounds of $n/2$ games are played. In each round, a
matching is performed over the teams by a simple algorithm, that considers
all circular permutations to generate the games. We give an illustration for 
$n=6$ teams, starting with the configuration:

\begin{equation*}
\begin{array}{lll}
1 & 2 & 3 \\ 
4 & 5 & 6.%
\end{array}%
\end{equation*}

This configuration implies that in the first round, team 1 plays team 4, 2
plays 5 and team 3 plays team 6. To generate the second round, we keep team 1 fixed in its position
and we rotate the other teams clockwise:

\begin{equation*}
\begin{array}{lll}
1 & 4 & 2 \\ 
5 & 6 & 3%
\end{array}%
\end{equation*}

Now, team 1 plays team 5, team 4 plays 6 and team 2 plays 3. After $n-1=5$
rounds, the system arrives at the last distinct configuration and all teams have confronted every 
other only once. We repeat the same process to simulate the second half of the tournament.

In our model, the outcome of each match is a draw with probability $r_{draw}$
and one team will beat the other with probability $(1-r_{draw})$; the
winning team is decided by the probabilities defined by Eq.(\ref{prob_win}).
After each match, we increase $\varphi_i $ by one unit if team $i$ wins,
decrease $\varphi_i $ by one unit if team $i$ loses and $\varphi_i \geq 1$, and
leave it unchanged in the case of a draw. Here, we used $r_{draw}=%
\allowbreak 0.26$ that is the average draw probability in actual tournaments
around the world. Actually, we observe that $r_{draw}$ ranges from $0.24$
(Spanish La Liga) to $0.28$ (Italian Calcio); see table~\ref{main_table}.

Besides this, the team is awarded points according to the 3-1-0 scheme. In
each new match, the updated potentials are considered and the second half of
the tournament begins with the conditions acquired by the teams in the first
half. The team evolution dynamics is briefly described by the following
algorithm:%
\begin{equation*}
\begin{tabular}{ll}
\hline\hline
\label{algorithm} & \textbf{Main} \textbf{Algorithm} \\ \hline\hline
1 & If ($rand[0,1]<r_{draw}$) then \\ 
2 & $\ \ \ \ \ p_{i}=p_{i}+1$ and $p_{j}=p_{j}+1;$ \\ 
3 & else \\ 
4 & \ \ \ if\ ($rand[0,1]<\frac{\varphi _{i}}{(\varphi
_{i}+\varphi _{j})}\ $) then \\ 
5 & \ \ \ \ \ \ \ \ \ $p_{i}=p_{i}+3;$\ \ $\varphi _{i}=\varphi _{i}+1;$\ $%
\varphi _{j}=\varphi _{j}-1;$\ \ \  \\ 
6 & \ \ \ else\ \ \ \ \ $\ $\ \  \\ 
7 & \ \ \ \ \ \ \ \ \ $p_{j}=p_{j}+3;$\ \ $\varphi _{i}=\varphi _{i}-1;$\ $%
\varphi _{j}=\varphi _{j}+1;$\  \\ 
8 & \ \ \ endif \\ 
9 & Endif \\ \hline\hline
\end{tabular}%
\ \ \ \ \ 
\end{equation*}%
Here, it is important to notice that the algorithm works under the
constraint $\varphi _{j}\geq 1$, for every $j$. %$\min\{\varphi_{j}\}=1$.
 It is important to mention that the arbitrary choice of increments equal to
one unit is irrelevant, since it is possible to alter the relative change in
potential by assigning it different starting values. For example, if  team $A
$\ is matched against team $B$\ in a certain round, we can denote by $N_{A}$\ and $N_{B}$%
\ the difference in number of wins and losses up to that round  ($N_{A}$\ and $N_{B}$\ can
be negative or postive integers) for each team. We can then write  $\varphi _{A}=\varphi _{0}+N_{A}$\ and $%
\varphi _{B}=\varphi _{0}+N_{B}$, so our model works with unitary
increments/decrements, i.e., $\Delta \varphi =1$. As can be observed, for arbitrary $\Delta
\varphi $\ we have invariance of probability:%
\begin{equation}
\begin{array}{lll}
\Pr (A\succ B) & = & \dfrac{\varphi _{A}}{(\varphi _{A}+\varphi _{B})} \\ 
&  &  \\ 
& = & \dfrac{\varphi _{0}\Delta \varphi +N_{A}\Delta \varphi }{\varphi
_{0}\Delta \varphi +N_{A}\Delta \varphi +\varphi _{0}\Delta \varphi
+N_{B}\Delta \varphi } \\ 
&  &  \\ 
& = & \dfrac{\widehat{\varphi }_{0}+N_{A}\Delta \varphi }{\widehat{\varphi }%
_{0}+N_{A}\Delta \varphi +\widehat{\varphi }_{0}+N_{B}\Delta \varphi }=%
\dfrac{\widehat{\varphi }_{A}}{\widehat{\varphi }_{A}+\widehat{\varphi }_{B}},
\end{array}
\label{invariance}
\end{equation}%
where $\widehat{\varphi }_{A}=\widehat{\varphi }_{0}+N_{A}\Delta \varphi $\
and $\widehat{\varphi }_{B}=\widehat{\varphi }_{0}+N_{B}\Delta \varphi $. 
 This simple calculation shows that we can start from an arbitrary
potential $\widehat{\varphi }_{0}$ for the players and have exactly the same results if we perform
increments according to $\Delta \varphi $. In this case our main algorithm must be changed to increment/decrement by $%
\Delta \varphi $\ instead of 1 and it is dependent on one parameter only, i. e., $\varphi_0/\Delta \varphi$. 

\section{Results Part I: Exploring the first model -- Calibrating parameters}

Before comparing our model with real data from tournaments played under the
DRRS, it is interesting to study some of its statistical properties. Given $%
n $ teams, one run of the algorithm will generate a final classification
score for each team. For instance, starting with $n=20$ teams with the same
potential $\varphi_{0}=30$, a possible final classification score generated
by our algorithm in increasing order is [23, 28, 39, 41, 44, 45, 47, 48, 49,
53, 54, 57, 60, 61, 62, 62, 64, 64, 65, 72]. To obtain significant
information from the model, it is necessary to average these data over
different random number sequences. To that end, we compute histograms of final
score distributions for $n_{run}=100$ different final scores, for a varying
number of teams.

\begin{figure}[h]
\begin{center}
\includegraphics[width=6.0in]{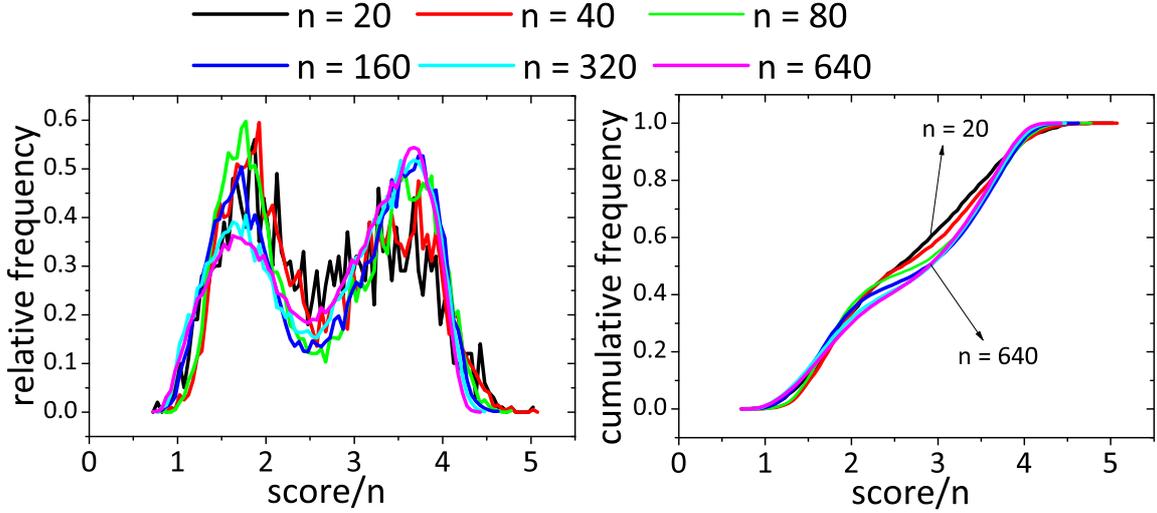}
\end{center}
\caption{ Figure (a): \textbf{The score histogram rescaled by the number of
teams in the large fluctuation regime tournament for} $\protect\varphi_{0}=2$. 
The histogram was generated from an average over $n_{run}=100$
different tournaments simulated by our automaton. We simulated tournaments
with $n=20,40,80,160,320$, and $640$ different teams. Figure (b): \textbf{The
accumulated frequency of classification scores}: number of teams
with score smaller than a determined score divided by the number of teams. }
\label{figure_scaling_2}
\end{figure}

In Fig.~\ref{figure_scaling_2} (a), we display the relative frequency of
scores as a function of the rescaled score, considering all teams initially
with $\varphi_{0}=2$, for varying tournament sizes $n$. Under this regime of
low $\varphi_0$, the changes in potential according to the algorithm generate
large fluctuations in the winning/losing probabilities and a double peak
pattern is observed in the histograms. 

For a study of scaling size, we consider our histogram as a function
of the variable $\frac{score}{n}$ since the larger the tournament the larger
are the team scores (number of points). This double peak shows that our
dynamics leads to two distinct groups: one that disputes the leadership and
the other that fights against relegation to lower tiers. In Fig.~\ref%
{figure_scaling_2} (b), we plot the cumulative frequency as a function of $\frac{score}{n}$
 (that essentially counts how many teams have scores smaller than, or equal to a
given score). We can observe an interesting behavior due to the presence of
extra inflection points that makes the concavity change sign and the 
non-gaussian behavior of the scores, independent of the size of the
tournaments. Although clearly non-gaussian, because of the double peak and
the ``S'' shaped cumulative frequency, the Kolmogorov-Smirnov (KS) and
Shapiro-Wilk(SW) tests (references and routine codes of these tests are found in 
\cite{recipes2007}) were performed to quantify the departure from
gaussianity. An important point for methods applied is KS \cite{Garpman1978} can be 
applied to test other distributions differently from SW which is used for normality 
tests specifically.  

\begin{figure}[h]
\begin{center}
\includegraphics[width=6.0in]{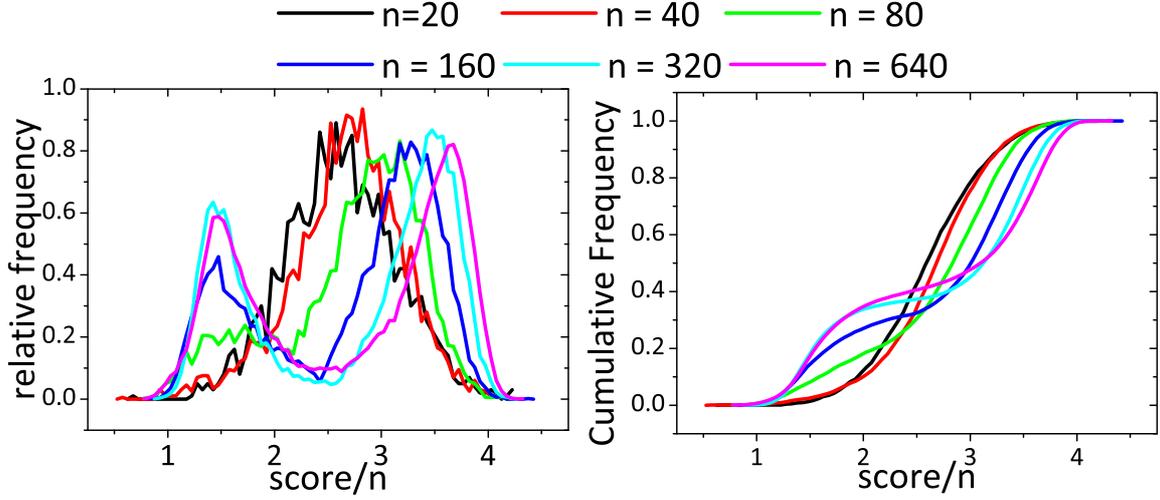}
\end{center}
\caption{Figure (a): \textbf{The histogram of scores rescaled by the number
of teams}: in the small fluctuation regime tournament for $\protect\varphi_{0}=30\ $,  
under the same conditions of Figure~\protect\ref{figure_scaling_2}. 
Figure (b): \textbf{The accumulated frequency for $\protect\varphi_{0}=30$}, for the 
same conditions of Figure~\protect\ref{figure_scaling_2}.}
\label{figure_scaling_30}
\end{figure}

By repeating the experiment for $\varphi_{0}=30$ a transition from a single
peak to a double peak can be observed from $n\approx40$ which is observed in
Fig. \ref{figure_scaling_30} (a). Under this condition, wins and losses cause
small changes in the winning/losing probabilities simulating a tournament
under the ``adiabatic" regime.

We observe that this interesting behavior is reflected in the curves of
cumulative frequencies that change from single to double ``S" shaped in Fig.~%
\ref{figure_scaling_30} (b). It is interesting to verify whether this
tournament model is able to mimic the score statistics of real tournaments
and, if so, under what conditions? To answer this question, we need to
explore real tournaments statistics. In Table~\ref{main_table}, we show the
compiled data of the last 6 editions of important soccer tournaments around
the world: Italian, Spanish, and Brazilian. We collect data about scores of
the champion teams (maximum) and last placed teams. We average these
statistics for all studied editions and we analyze the Gaussian behavior of
score data for each edition separately (20 scores) and grouped (120 scores)
by using two methods: Shapiro-Wilk and Kolmogorov-Smirnov using a
significance level of 5\%. The draw average per team was also computed,
which shows that $r_{draw}\approx\frac{10}{38} \approx 0.26$ which corroborates the
input used in our previous algorithm.

\begin{table*}[th]
\caption{ \textbf{Compiled data of important tournaments around the world:
Italian, Spanish, and Brazilian Leagues}}
\label{main_table}%
\begin{tabular}{llllllll}
\hline\hline
& \textbf{2006} & \textbf{2007} & \textbf{2008} & \textbf{2009} & \textbf{%
2010} & \textbf{2011} & \textbf{all} \\ \hline\hline
\textbf{Italian* (Calcio)} &  &  &  &  &  &  &  \\ \hline
minimum & 35 & 26 & 30 & 30 & 29 & 24 & 29(2) \\ 
maximum & 86 & 97 & 85 & 84 & 82 & 82 & 86(2) \\ 
Kolmogorov-Smirnov & no & yes & yes & yes & yes & yes & no \\ 
Shapiro-Wilk & no & no & yes & yes & yes & yes & no \\ 
draws (average per team) & 12.5 & 11.4 & 11.2 & 9.5 & 10.2 & 9.7 & 10.8(5)
\\ \hline\hline
\textbf{Spanish (La Liga)} &  &  &  &  &  &  &  \\ \hline
minimum & 24 & 28 & 26 & 33 & 34 & 30 & 29(2) \\ 
maximum & 82 & 76 & 85 & 87 & 99 & 96 & 87(4) \\ 
Kolmogorov-Smirnov & yes & yes & yes & yes & yes & yes & no \\ 
Shapiro-Wilk & yes & yes & yes & yes & no & no & no \\ 
draws (average per team) & 10.5 & 9.8 & 8.7 & 8.3 & 9.5 & 7.9 & 9.1(4) \\ 
\hline\hline
\textbf{Brazilian (Brasileir\~{a}o)} &  &  &  &  &  &  &  \\ \hline
minimum & 28 & 17 & 35 & 31 & 28 & 31 & 28(2) \\ 
maximum & 78 & 77 & 75 & 67 & 71 & 71 & 73(2) \\ 
Kolmogorov-Smirnov & yes & yes & yes & yes & yes & yes & yes \\ 
Shapiro-Wilk & yes & no & yes & yes & yes & yes & yes \\ 
draws (average per team) & 9.7 & 9 & 9.6 & 10.2 & 11.8 & 10.5 & 10.1(4) \\ 
\hline\hline
\end{tabular}%
\par
\begin{flushleft}
* The 2006 year (which corresponds to season 2005/2006) was replaced by
2004/2005 in Calcio since cases of corruption among referees have led to
changes in teams scores with points being reduced from some teams and
assigned to others. Here ``yes" denotes positive to normality test and ``no"
denotes the opposite, at a level of significance of 5\%.
\end{flushleft}
\end{table*}

Some observations about this table are useful. The traditional European
tournaments, based on the DRRS have non-Gaussian traces as opposed to the
Brazilian league, an embryonary tournament played under this system. This
fact deserves some analysis: in Brazil, over the last 6 editions,
(compiled data are presented in Table \ref{main_table}) 4 different football clubs
have won the league. If we consider all 10 disputed editions, we have 6 different
champions which shows the great diversity of this competition. The Brazilian
League seems to be at a greater random level when compared to the European
tournaments. A similarity among teams suggests that favorites are not always
crowned champions and many factors 
and small fluctuations can be decisive in the determination of the champion.
This may also indicate that the Brazilian tournament has an abundance of
homogeneous players differently from the Italian tournament, in which the
traditional teams are able to hire the best players or have well-managed
youth teams, or even sign the ones who play for the national Italian team.
Consider for example Real Madrid and Barcelona in Spain: they govern the
tournament by signing the best players, even from youth teams from abroad (as
is the case with the World's Player of the Year, Lionel Messi who joined
Barcelona at age 13 from Argentina). It is not uncommon for a player who has stood
 out in the Brazilian or other latin american champioships to be hired to play in 
Europe for the next season, further contributing to the lack of continuity from one 
season to the next and to the ``randomization'' of the teams. 

In Brazil, there is not a very large financial or economic gap among teams
and although favorites are frequently pointed out by sports pundits before
the beginning of the tournament, they are typically not able to pick the
winners beforehand. In fact, many dark horses, not initially pointed out as
favorites, end up winning the league title. This suggests that, in Brazil,
the champions emerge from very noisy scenarios, as opposed to other
tournaments that only confirm the power of a (favorite) team. One could add
to that the existence of a half-season long local (or state-based only)
tournament making the predictions widely reported in the press not very
trustful or reliable in any sense. Therefore, it is interesting to check our
model by studying its statistical properties, changing parameters and then
comparing the model with real data.

\begin{figure}[h]
\begin{center}
\includegraphics[width=3.5in]{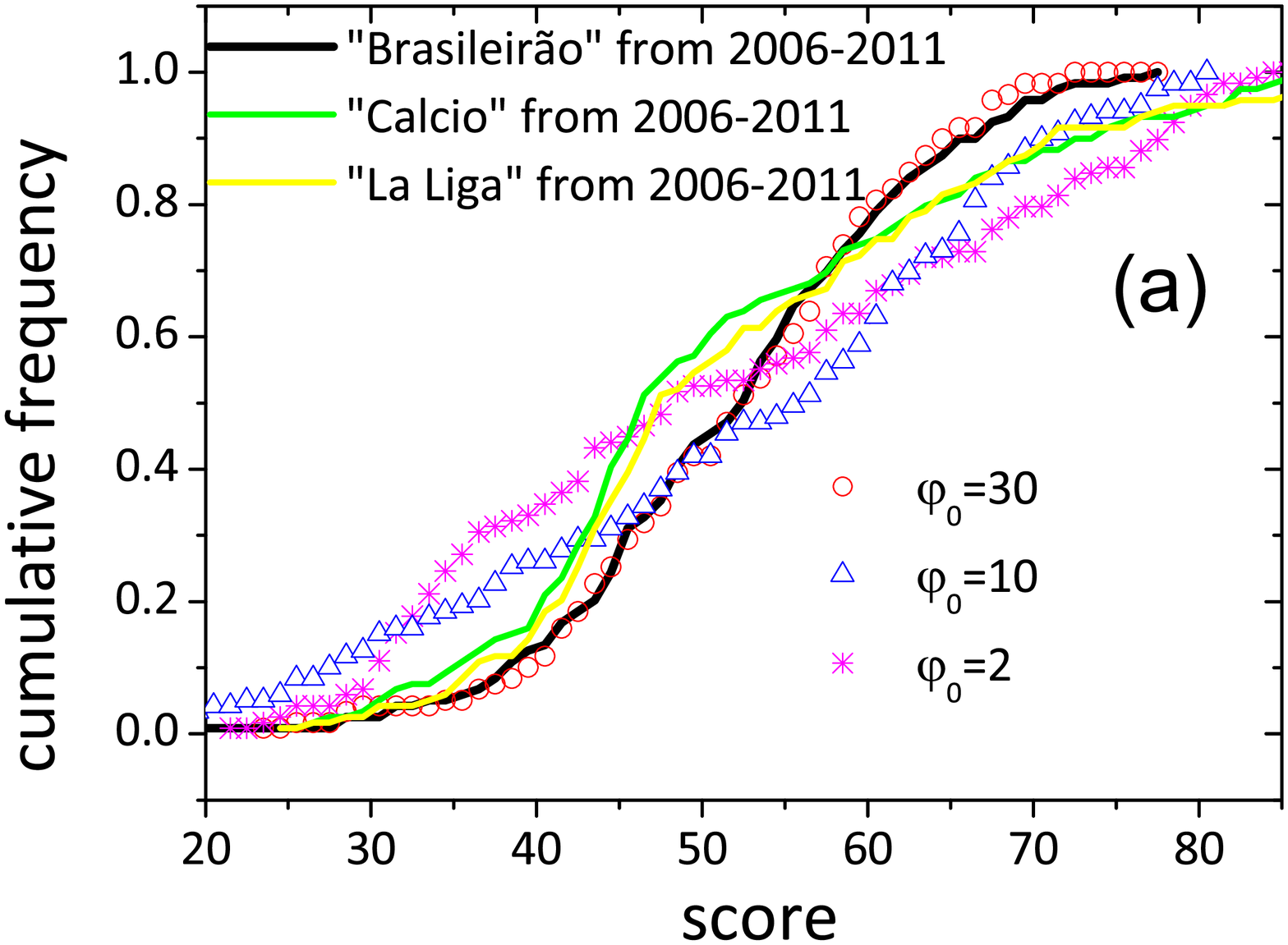}\includegraphics[width=3.5in]{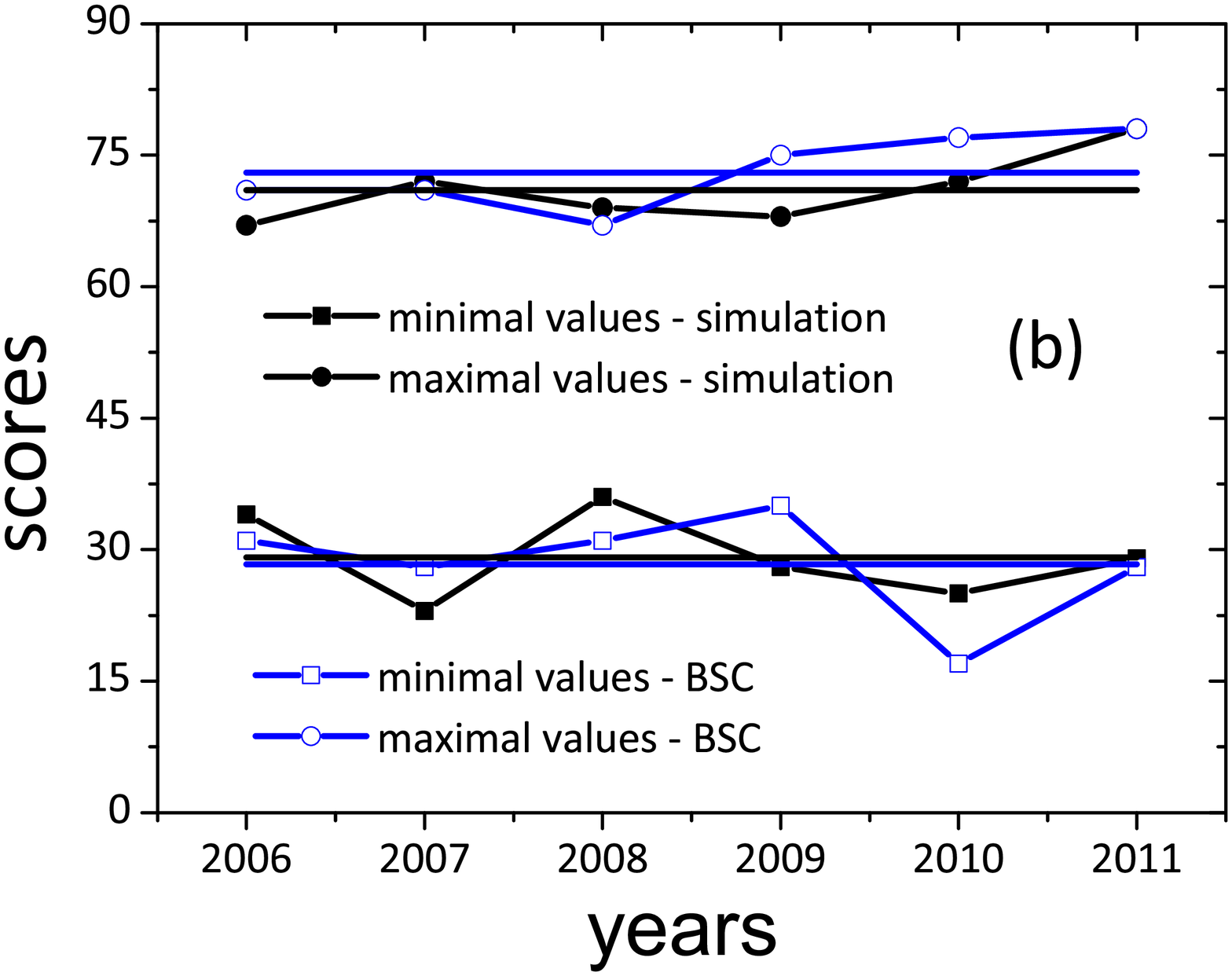}
\end{center}
\caption{Figure (a):\textbf{ A comparison between results produced
by our model, using different initial parameters $\protect\varphi_{0}=2,10$ 
and $30$ and different tournaments}. We have used $r_{draw}=0.26$. 
Results were obtained considering 6 runs of ourartificial tournament. We 
can observe that the model fits the Brazilianleague (black continuous curve 
obtained from 6 editions of the Brazilian league) precisely for $\protect\varphi_{0}=30$. 
On the other hand, Calcio and La Liga are not reproduced by our model indicating clear
differences between such tournaments and the Brasileir\~{a}o. Figure (b): 
\textbf{This figure shows that minimal and maximal values obtained by our model} 
(full squares and circles respectively, depicted in black) are very similarto the ones 
obtained in the six editions of the Brazilian League (open
squares and circles, in blue). The continuous line corresponds to the
average values obtained in each case for a comparison.}
\label{results_cumulative}
\end{figure}

We perform an analysis of our model considering different initial parameters 
$\varphi _{0}=2,10$ and $30$ and the different tournaments (see plot (a) in
Fig. \ref{results_cumulative}). We have used $r_{draw}=0.26$ in our
simulations. It is possible to observe that the model fits the Brazilian
soccer very well for $\varphi _{0}=30$ and $\Delta \varphi =1$. It is
important to mention that extreme values (minimal and maximal) are
reproduced with very good agreement. For example, plot (b) in Fig. \ref%
{results_cumulative} shows that minimal and maximal values obtained by our
model (full squares and circles respectively, in black) are very similar to
the ones obtained from the six editions of the Brazilian tournament (open
squares and circles, in blue). We also plot continuous lines that represent
 the average values obtained in each case. This shows
that our model and its fluctuations capture the nuances and emerging
statistical properties of the Brazilian tournament which, however, seems not
to be the case of Calcio and La Liga. Plot (a) of Fig. \ref%
{results_cumulative} shows that the cumulative frequency of these two
tournaments are very similar to one another and that no value of the
parameter $\varphi _{0}$ (many others were tested) 
is capable of reproducing their data.

Now a question that can quickly come to mind to readers of this paper is:
are we modeling something that is entirely random and non-evolutionary,
i.e., could we use a simpler model? The answer, fortunately, is ``not
really''. To understand this, let us suppose a completely random and non
evolutionary model (the probabilities do not change with time), in which a
team should win, lose, or draw with the same probability: 1/3.

\begin{figure}[h]
\begin{center}
\includegraphics[width=3.5in]{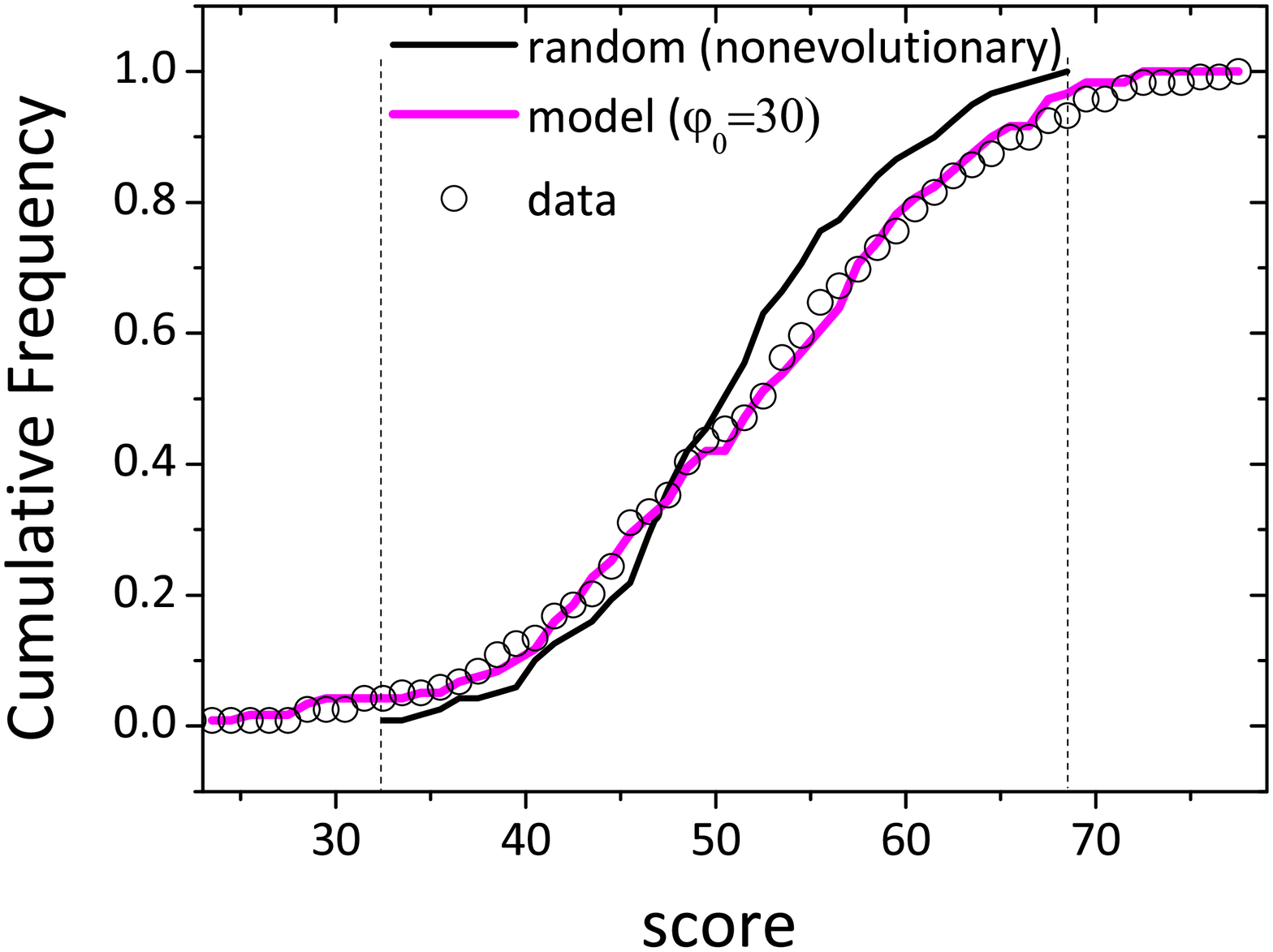}
\end{center}
\caption{\textbf{Comparison of our model (evolutionary) and a totally random
(static) model}: We can observe that the static model does not reproduce the
lower and upper values as well as the shape of the Brazilian League which,
on the other hand, is very well fitted by our model. }
\label{evolutionary_vs_non_evolutionary}
\end{figure}

A comparison of the best fit of our model (evolutionary) with the totally
random (static) model, under the exact same conditions of 20 teams under the
DRRS, is shown in Fig. \ref{evolutionary_vs_non_evolutionary}. We observe
that the static model does not reproduce the lower and upper values as well
as the shape of cumulative frequency of the Brasileir\~{a}o which, on the other hand, is very well
fitted by our model.

\section{Second model: Draw probabilities emerging from the model itself}

Previous authors (see for example \cite{Bitner2007} and \cite{Skinera2009})
claim that in a match between two teams $A$ and $B$, the probability that
the result is ($n_{A}$, $n_{B}$), where $n_{i}$ is the number of goals
scored by team $i$, can be approximated by a Poisson distribution:%
\begin{equation}
\begin{array}{lll}
\Pr \left[ (n_{a},n_{b})|(\phi _{A},\phi _{B})\right] & = & \Pr (n_{a},\phi
_{A})\cdot \Pr (n_{b},\phi _{B}) \\ 
&  &  \\ 
& = & \frac{\phi _{A}^{n_{A}}}{n_{A}!}e^{-\phi _{A}}\cdot \frac{\phi
_{B}^{n_{B}}}{n_{B}!}e^{-\phi _{B}}%
\end{array}
\label{poisson}
\end{equation}%
where $\phi _{A}$ and $\phi _{B}$, the average number of goals in a game,
are taken as the abilities of the teams.

\begin{figure}[h]
\begin{center}
\includegraphics[width=3in]{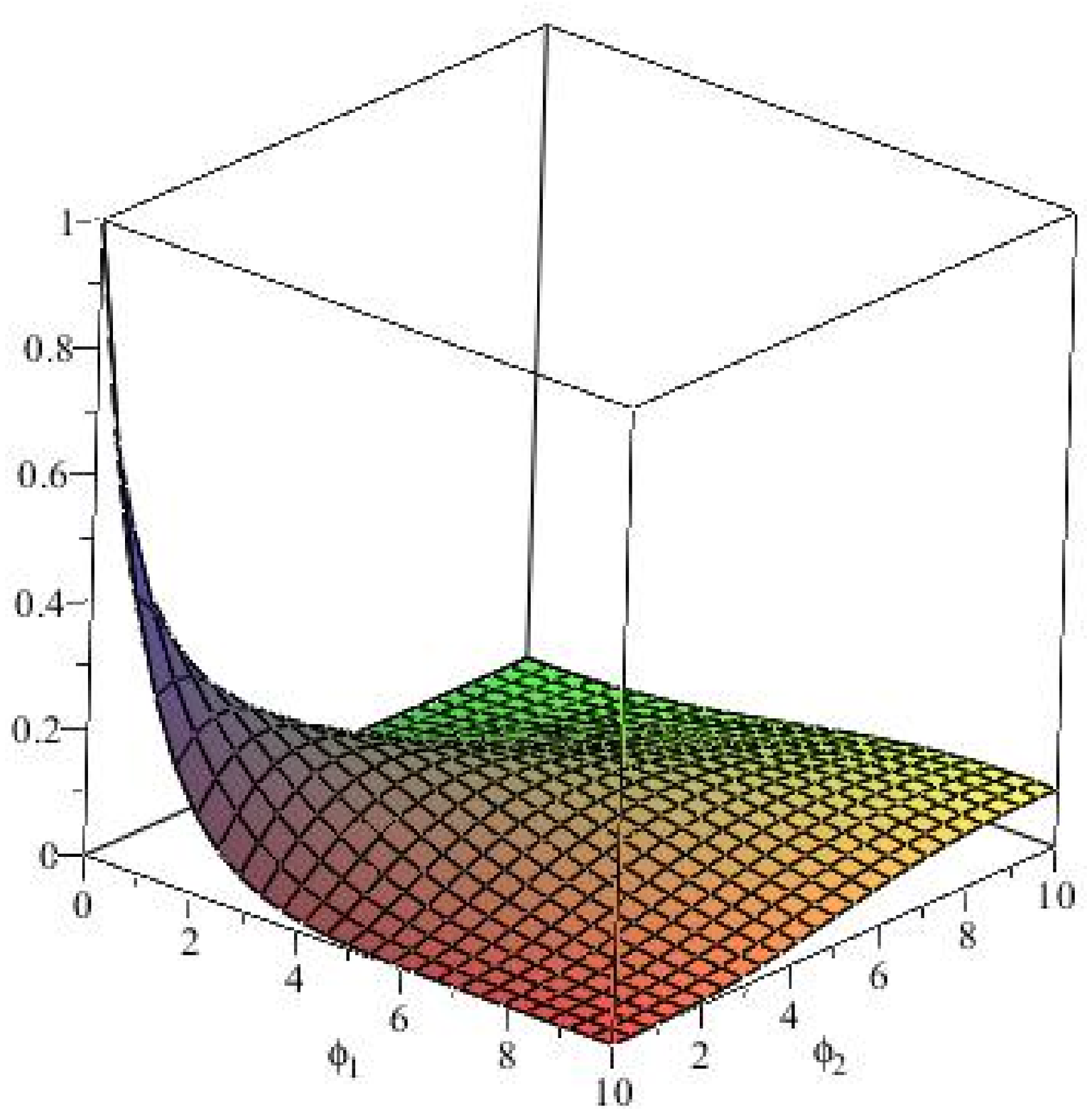}
\end{center}
\caption{The probability $r_{draw}$ as function of $\protect\phi_{A}$ and $\protect\phi_{B}$%
. }
\label{plot_m_infinity}
\end{figure}

It is very interesting to work with models that have as few free parameters
as possible; the imposition of an ad-hoc draw probability in our previous
version of the model can therefore be seen as a shortcoming. It is possible
to overcome this problem, and at the same time maintain the same model
properties, by using Eq.~\ref{poisson} as a means to calculate $r_{draw}$ in
the previously defined algorithm. Given two teams, with potentials $%
\varphi_{A}$ and $\varphi_{B}$, we can calculate $r_{draw}$ making the
direct identification of the abilities with our concept of potential, i.e., $%
\phi_{A}=\varphi_{A}$ and $\phi_{B}=\varphi_{B}$, so that 
\begin{equation*}
\begin{array}{lll}
r_{draw} & = & \Pr\left[ (n_{a}=n_{b})|(\phi_{A},\phi_{B})\right] \\ 
&  &  \\ 
& = & {\displaystyle\sum\limits_{n=0}^{\infty}} \frac{\phi_{A}^{n}}{n!}%
e^{-\phi_{A}}\cdot\frac{\phi_{B}^{n}}{n!}e^{-\phi_{B} } \\ 
&  &  \\ 
& = & {\displaystyle\sum\limits_{n=0}^{\infty}} \frac{(\phi_{A}\phi_{B})^{n}%
}{n!^{2}}e^{-(\phi_{A}+\phi_{B})},%
\end{array}%
\end{equation*}
leaving our model with only one free parameter.

The first important point is that the probability independent of previous
ad-hoc information obtained from tournaments, arising as a property of the
teams themselves, i.e., their ability to score goals. A plot of $r_{draw}$ as function of $\phi _{A}$ and $\phi
_{B}$ is shown in fig.~\ref{plot_m_infinity}. However it is important to
mention this definition must be adapted if $\varphi $ is not exactly the
average number of goals of the team per match ($\phi $), since the number of
goals in a match is finite, its extension to infinity can have drastic
consequences in the draw probabilities. It should be noted that the
potential of the teams (which may be rescaled) represents the abilities of
teams but can be very different from the average of the number of goals
scored by teams in a given match. In this case, a solution to this problem
is to consider a truncated Poisson function%
\begin{equation*}
f^{trunc}(n,\phi )=Z(\phi ,m)^{-1}\frac{\phi ^{n}e^{-\phi }}{n!}
\end{equation*}%
where 
\begin{equation*}
Z(\phi ,m)={\displaystyle\sum\limits_{j=0}^{m}}\frac{\phi ^{j}e^{-\phi }}{j!}
\end{equation*}%
with $m$ being the appropriate cutoff for modeling and must be suitably
adjusted. Therefore, $r_{draw}$ is now re-written as%
\begin{equation}
r_{draw}=\frac{Z(\phi _{B},m)^{-1}}{Z(\phi _{A},m)}{\displaystyle%
\sum\limits_{n=0}^{m}}\frac{(\phi _{A}\phi _{B})^{n}}{n!^{2}}e^{-(\phi
_{A}+\phi _{B})}\text{.}  \label{new_rdraw}
\end{equation}

This is a solution but $m$\ must be adjusted according to the initial potential $%
\varphi _{0}$\  if we use $\Delta \varphi =1$. 
 However, it is also possible to solve this problem by suitably scaling the potential to be 
the average number of goals in a match. Therefore, if we start the simulations with the potential 
representing the average number of goals of a real tournament $\widehat{\varphi }_{0}=\lambda =$\ $\varphi
_{0}\Delta \varphi $, then the increment must be given by  $\Delta \varphi =\lambda
/\varphi _{0}$, so that the the win/lose probabilities are kept fixed, according to
equation \ref{invariance}. In this case, $m\rightarrow \infty $ presents the best fits and gives the correct
 draw probabilities $r_{draw}$, making the model again more
suitable, since $m$\ is not an experimental parameter.\textbf{\ }

In the next section, we will present our results based on this new approach 
for the calculation of $r_{draw}$ and we show that real data are also reproduced by both methods
presented in this section. 

\section{Results Part II: variable draw probabilities}

We perform new simulations considering our previously described algorithm,
but allowing for variable draw probabilities. As before, we organize the
teams via the DRRS, starting with fixed potentials and take averages over
many runs of the model. Our first analysis was to reproduce the final score
of the Brazilian tournament tuning different values of $m$.

\begin{figure}[h]
\begin{center}
\includegraphics[width=2.0in]{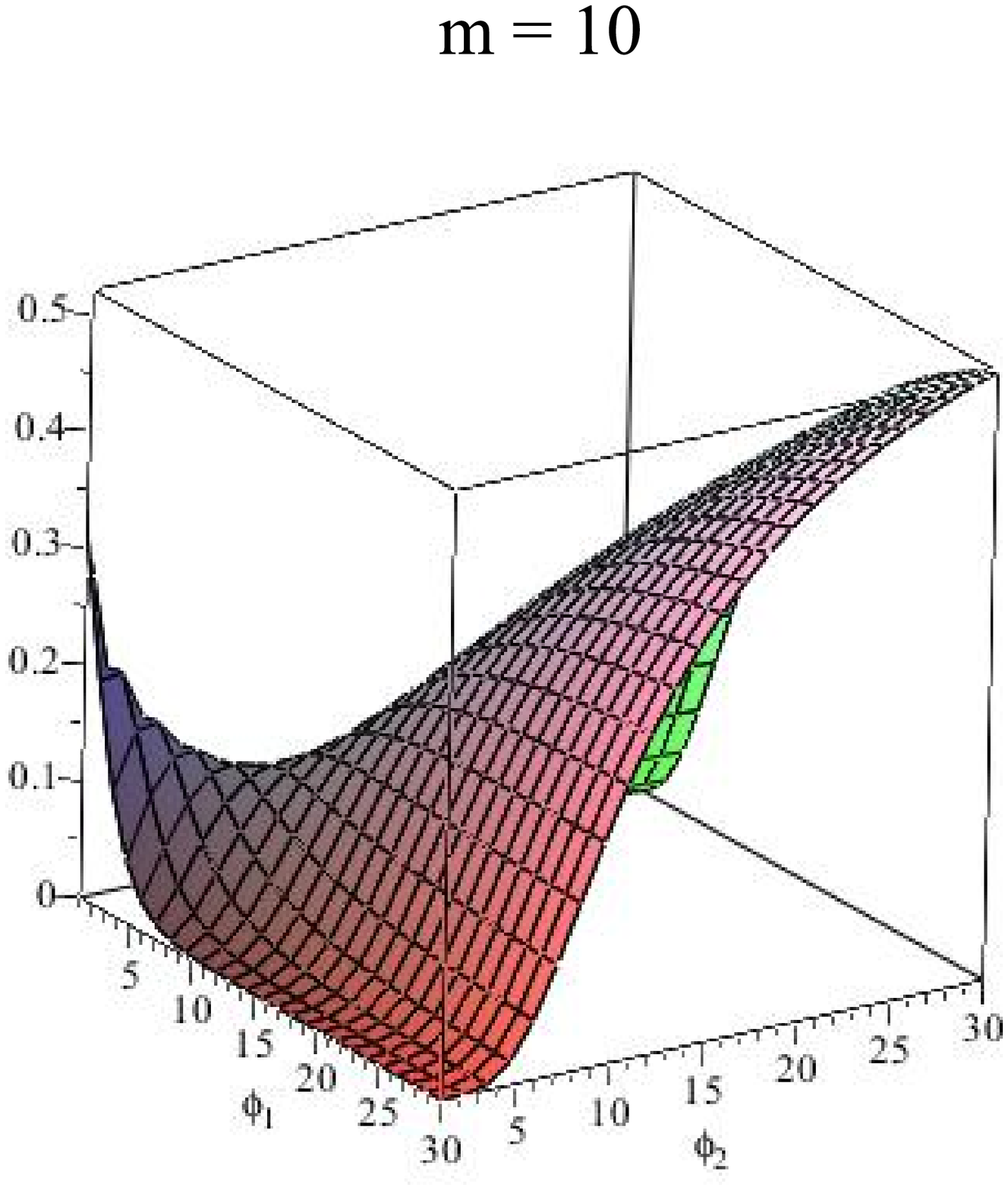} %
\includegraphics[width=2.0in]{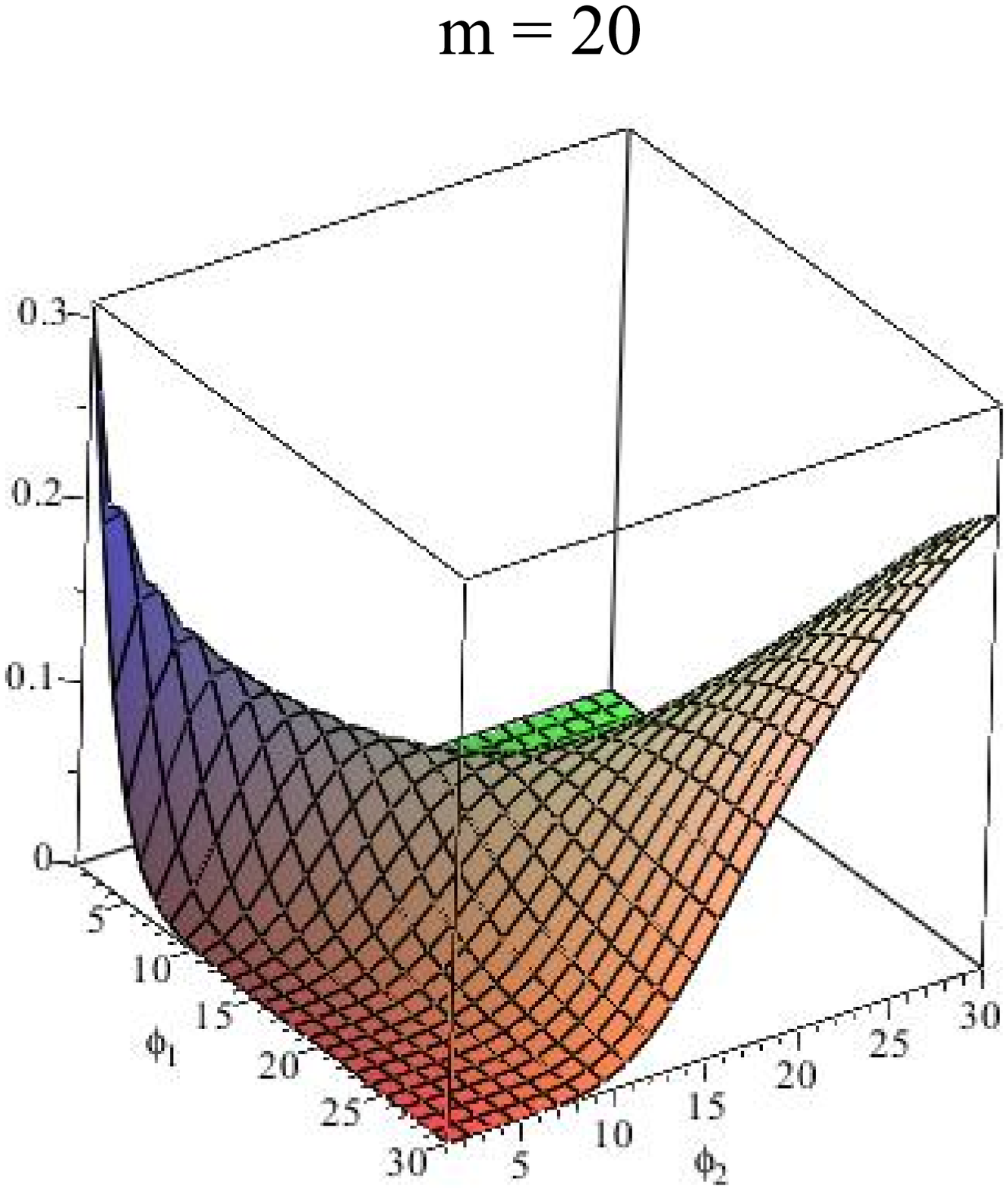} %
\includegraphics[width=2.0in]{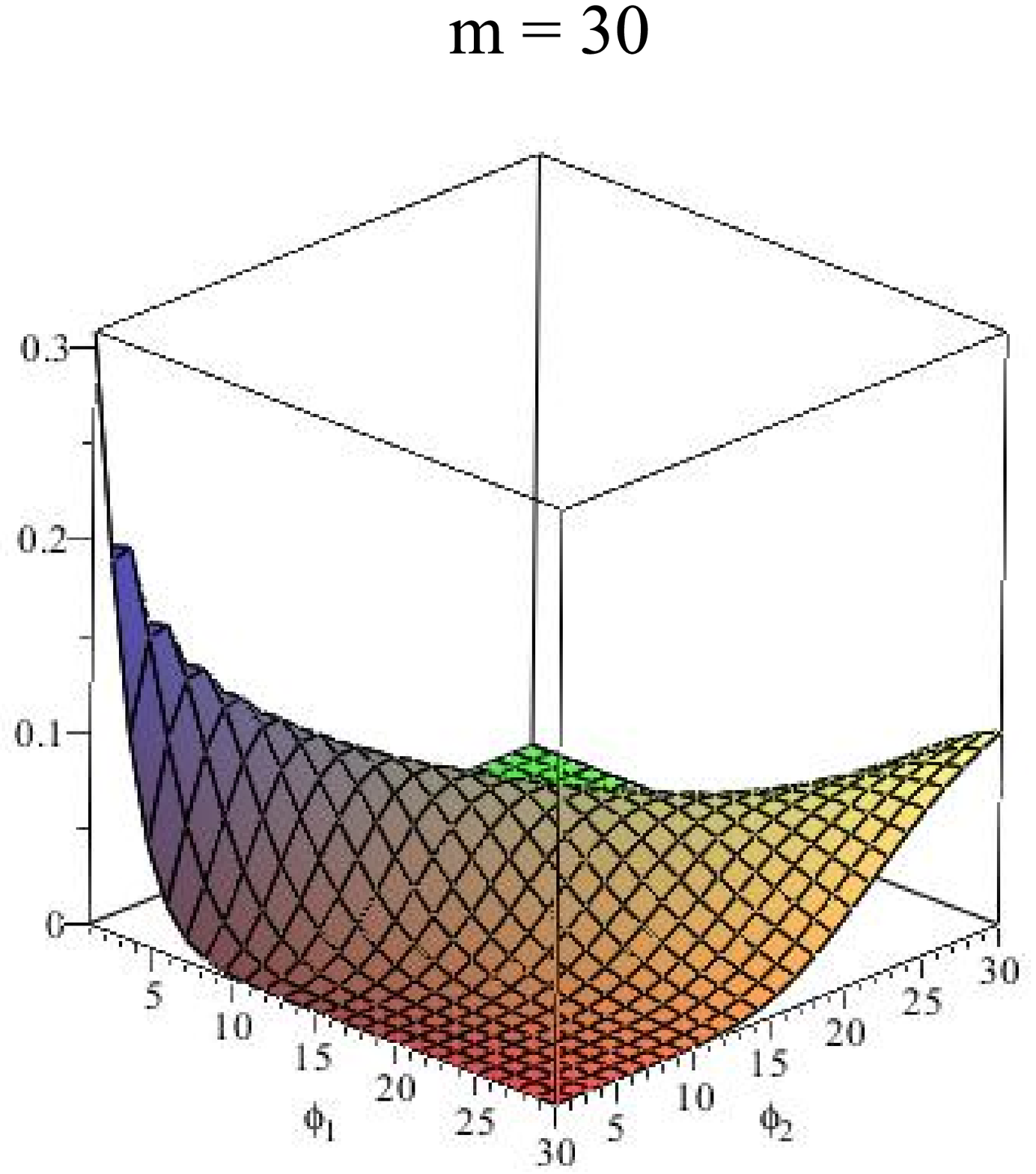}
\end{center}
\caption{Plots of $r_{draw}$ given by equation \protect\ref{new_rdraw} for
different values of $m$.}
\label{different_m}
\end{figure}

Figure \ref{different_m} shows the surfaces corresponding to $r_{draw}$
calculated by equation~\ref{new_rdraw}. We can see that higher $m$ values
result in higher draw probabilities for low potentials. Figure \ref%
{cumulative_different_m} shows the cumulative distribution of simulated
final scores from our artificial tournament generated by the model
considering the variable $r_{draw}$ given by equation \ref{new_rdraw}. Three
different values of $m$ (10, 20 and 30) were tested. We can show that best
value to fit the real data extracted from the same 6 Brazilian soccer
tournaments (continuous curve) is $m=20$. All teams started the simulations
with$\ \varphi _{0}=\phi _{0}=30$, calibrated in the previous results
developed in Results Part I.

\begin{figure}[h]
\begin{center}
\includegraphics[width=3.5in]{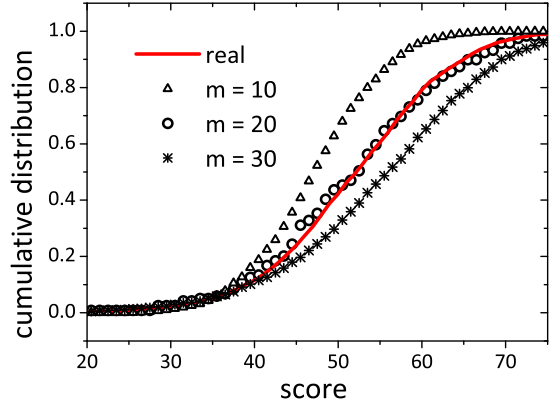}
\end{center}
\caption{Cumulative distribution of simulated tournament generated by the
model considering the variable $r_{draw}$ given by equation \protect\ref%
{new_rdraw}. We show that the best value to fit the real data extracted
from the same 6 Brazilian soccer tournaments (continuous curve) is $m=20$. }
\label{cumulative_different_m}
\end{figure}

\begin{figure}[h]
\begin{center}
\includegraphics[width=3.5in]{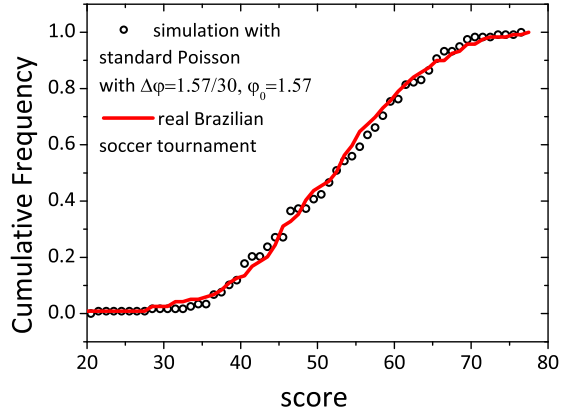}
\end{center}
\caption{Cumulative distribution of the simulated tournament generated by the model, 
considering $r_{draw}$ calculated from the standard Poisson distribuition ($%
m\rightarrow \infty $) (black balls) averaged over 6 repetitions. The continous
curve shows the 6 editions of the Brasilian soccer tournament. }
\label{standardt_poisson}
\end{figure}

Since $m$\ is not an acessible parameter of tournaments, we can start from $%
\widehat{\varphi }_{0}=1.57$\ as the initial potential of the teams, which corresponds to the
average number of goals scored by a team in a match of the Brazilian tournament studied, and ajust 
$\Delta \varphi =1.57/30$ in our algorithm. In this case, an 
excelent fit (see figure \ref{standardt_poisson}) is obtained, considering  the regular Poisson 
distribution ($m\rightarrow \infty $). Naturally the
number 30 follows from our initial calibration of our model (when we fixed $%
r_{draw}=0.26$, and $\varphi _{0}=30$\ led to an excelent fit as we
previously observed). 

As can be seen in the figures, we can obtain good fits with this new version
of model, which is a little more complicated than the previous version with
constant drawing probabilities, but it uses information inherent in the
model itself.

\begin{figure}[h]
\begin{center}
\includegraphics[width=3.5in]{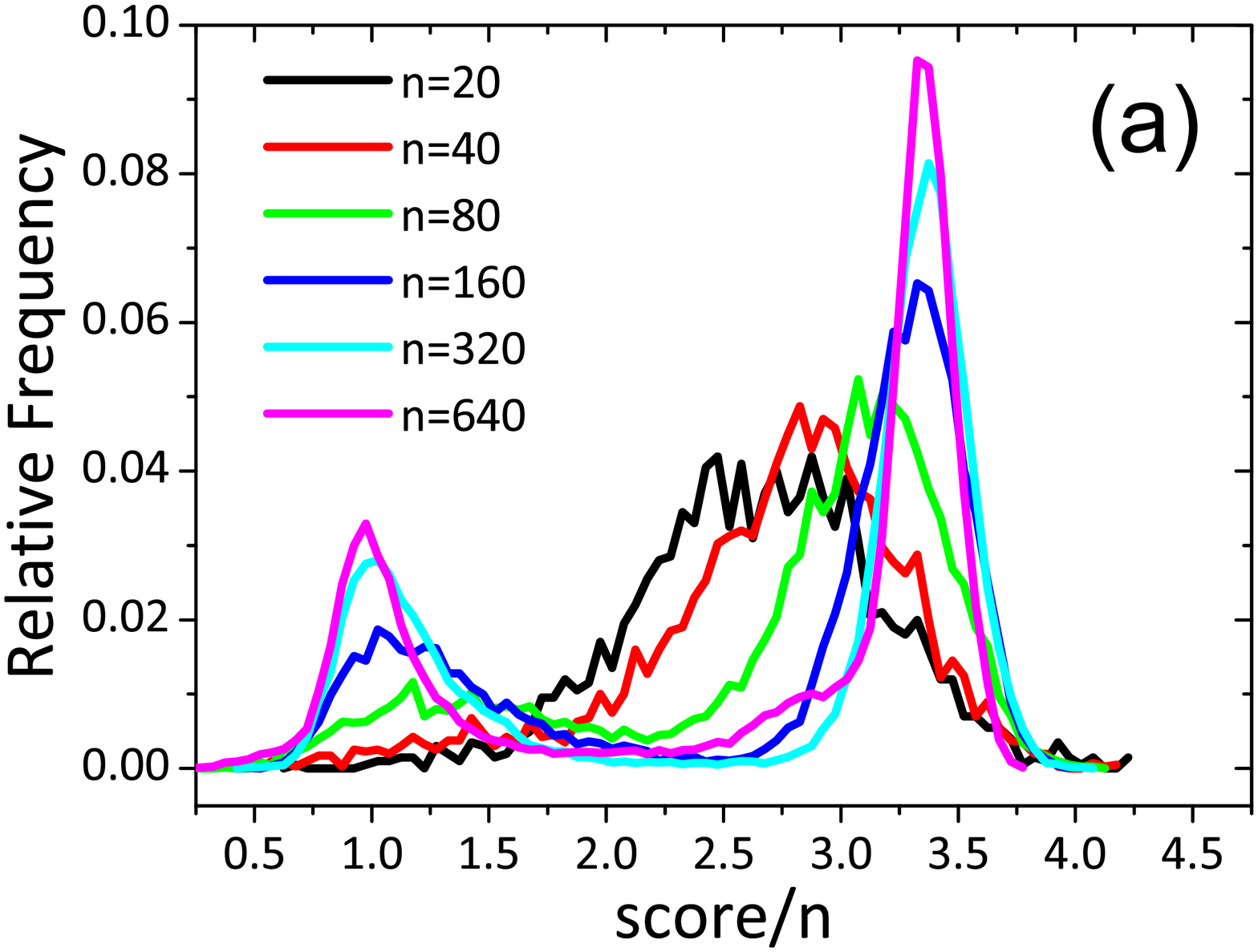}\includegraphics[width=3.5in]{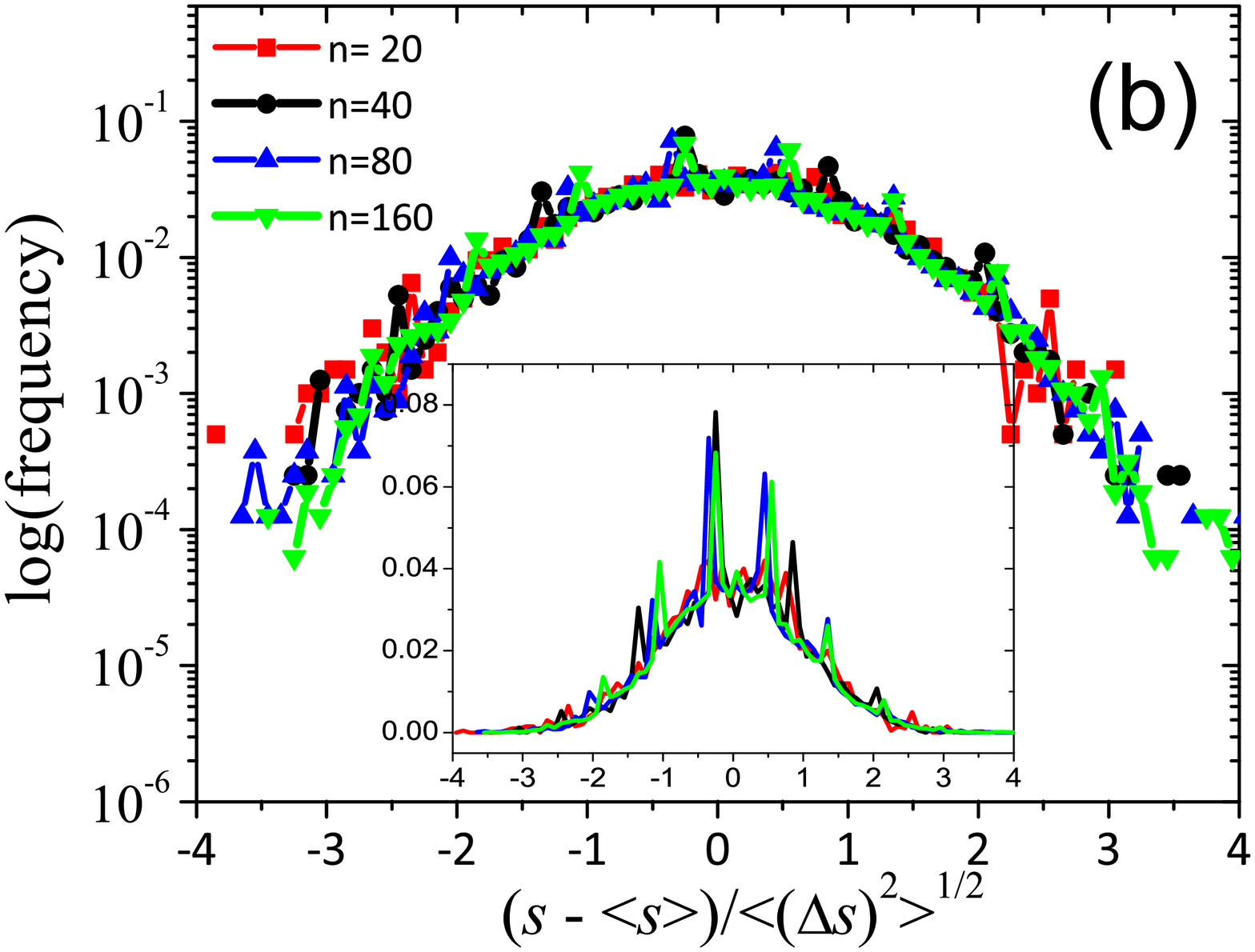}
\end{center}
\caption{\textbf{Plot(a)}: Histograms of final scores generated with $%
n_{run}=100$ different simulated tournaments using variable $r_{draw}$. The
figure is similar to figure \protect\ref{figure_scaling_30} since the same $%
\protect\varphi_{0}=30$ was used. \textbf{Plot (b)}: Scaling corresponding
to plot (a).}
\label{scaling}
\end{figure}

Finally, to test some scaling properties of the model, we reproduce the same plot of 
figure \ref{figure_scaling_30} using 
$r_{draw}$ according to \ref{new_rdraw}, which is shown in plot (a) in
figure \ref{scaling}.
This figure shows histograms of final scores generated by $n_{run}=100$
different simulated tournaments using variable probabilities $r_{draw}$. We can observe that 
the transition from one to two peaks is fully
due to the imposition that a team in our model has a minimal potential $%
\varphi \geq_i 1$. This effect can be overcome if we scale $\varphi_{0}$  with size system and, 
it is possible to collapse the curves
by re-writing the scores as normal standard variables, i.e., 
\begin{equation*}
score(n)=s(n)\rightarrow s^{\ast}(n)=\frac{s(n)-\left\langle
s(n)\right\rangle }{<(\Delta s(n))^{2}>}.
\end{equation*}
Thus, if $H(s^{\ast}(n),\varphi_{0},n)$ denotes the histogram of normalized
scores, we have the scaling relation $H(s^{\ast}(n_{1}),%
\varphi_{0},n_{1})=H(s^{\ast}(bn_{1}),b\varphi_{0},bn_{1})$. Plot (b) in
figure \ref{scaling} shows this scaling. We take the logarithms of the
histogram to show the collapse more explicitly. The small inset plot is
taken without the logarithm. We can see that different tournaments can
present the same properties as long as the teams' potentials are rescaled.
\

\section*{Summary and Conclusions}

In this paper, we have explored a new model that reproduces in detail the
final classification score (standings) of the teams in the Brazilian Soccer
tournament. The Brazilian tournament, as opposed to other tournaments such
as the Italian and Spanish Leagues, has some peculiarities and seems to
display scores that emerge from a dynamics that preserves its Gaussian
traces. This can be justified by several reasons: Brazilian tournaments have
many distinct champions and the competition is not dominated by a few teams.
Favorite teams frequently perform badly, and there is an inexhaustible
source of new players, making the tournament more balanced and very
susceptible to small fluctuations. Our model also seems to be a good
laboratory to study fluctuations that may happen in large tournaments. More
specifically the model presents a transition from a one to a two peaked
distribution of the final scores (standings) histograms that correspond to
disputes near the champion's region and another closer to the region of the
last placed team. Moreover, we also presented results relative to scaling of
histograms of final scores and showed that tournaments based in our model
for different sizes collapse on the same curve when we consider normal standard
deviations for final scores and a linear scaling for potentials.

Here, it is important to mention that after the present contribution 
was completed, we were alerted of the existence of a similar model with more parameters 
proposed to study statistical properties of tournaments~\cite{ribeiro2010}. However, our contributions is very different,
because in that study, the matches are generated under the mean field approximation regime based on a 
Markovian random walk. In such an approximation, therefore, the teams do not evolve in time.

\section*{About data extraction for the validation of the model}

Table \ref{main_table} shows the data from real tournaments used to compare
with the results produced by our model, as illustrated in Figs.~\ref%
{results_cumulative} and \ref{evolutionary_vs_non_evolutionary}. The data
(available publicly at http://www.wikipedia.org/) is based on records from
the Italian, Spanish, and Brazilian tournaments during the 2005/2006 -
2010/2011 seasons. For the Italian Calcio, the year 2006 (which corresponds
to season 2005/2006) was replaced by 2004/2005, since cases of corruption
and a referee scandal in 2005/2006 have supposedly changed the scores of
teams, and points were reduced from some teams and awarded to others. To
obtain the data from our model, we implemented a simple algorithm in the
FORTRAN language which computes the possible games according to the DRRS
system and evolves the potential and points of teams producing a final
classification score, or even a large number of final classification scores.
This is used to plot Figures \ref{figure_scaling_2} and \ref%
{figure_scaling_30} which explore the details of the model. All other
figures of the paper represent a comparison of the data extracted from
Wikipedia and those produced by our model.

\section*{Acknowledgments}

The authors thank CNPq (The Brazilian National Research Council) for its
financial support.

\bibliography{2012_05_june_soccer.bbl}

\end{document}